\documentclass[11pt,a4paper]{article}
\usepackage{jheppub}
\usepackage{color,subfigure}

\title{Quark-antiquark Scattering Phase Shift and Meson Spectral Function in Pion Superfluid}
\author{Tao Xia$^{1}$, Jin Hu$^{2}$ and Shijun Mao$^{2}$
       }
\affiliation[]{$^{1}$ College of Optoelectronic Science and Engineering, National University of Defense Technology, Changsha 410073, P. R. China\\
$^{2}$School of Science, Xi'an Jiaotong University, Xi'an 710049, P. R. China}
\emailAdd{maoshijun@mail.xjtu.edu.cn}

\abstract{We study the quark-antiquark scattering phase shift and meson spectral function in the pion superfluid described by the Nambu--Jona-Lasinio model. The meson mixing in the pion superfluid changes dramatically the full scattering phase shift and spreads strongly the spectral function of some of the collective modes.
}

\begin{document}
\maketitle

\section{Introduction}
The study on the phase structure of Quantum Chromodynamics (QCD) has been extended from finite temperature and baryon density to finite
isospin density. The physical motivation to study QCD at
finite isospin density and the corresponding pion superfluidity
is related to the investigation of compact stars,
isospin asymmetric nuclear matter, and heavy ion collisions
at intermediate energies. In early studies on dense
nuclear matter and compact stars, it has been suggested
that charged pions and even kaons are condensed at sufficiently
high density~\cite{migdal,schmitt}.

It is in principle no problem to do lattice simulation at finite
isospin density~\cite{bazavov}. It has been found that~\cite{kogut1,detmold} there is a phase transition from normal phase to pion
superfluidity at a critical isospin chemical potential which
is about the pion mass in vacuum, $\mu_I\simeq M_\pi$. The QCD
phase structure at finite isospin density is also investigated
in many low energy effective models, such as chiral perturbation
theory~\cite{son,kogut2,loewe}, Nambu--Jona-Lasinio (NJL)
model~\cite{toublan,barducci1,he1,ebert,abuki,andersen,xia}, random matrix method~\cite{klein,arai}, ladder
QCD~\cite{barducci2}, and strong coupling lattice QCD~\cite{nishida}.

One of the models that enables us to see directly how the
dynamic mechanisms of spontaneous symmetry breaking and
restoration operate is the NJL model~\cite{njl} applied to quarks~\cite{vogl,klevansky,volkov,hatsuda,buballa}.
In this model, quarks are elementary particles and mesons are quantum fluctuations. At mean field level to quarks
and in random phase approximation (RPA) to mesons, one can express the fluctuation contribution to the thermodynamics of the quark-meson system in terms of the bound states and scattering phase shifts of quark-antiquark pairs~\cite{hufner,zhuang}. This result appears to be more general than for just the NJL model: it resembles the Beth-Uhlenbeck formula for the second virial coefficient for a gas of non-relativistic particles~\cite{beth,uhlenbeck}, and its relativistic generalization by Dashen et al.~\cite{dashen}. A non-relativistic approach including medium effects in a nuclear physics content was used by Schmidt et al.~\cite{schmidt}. Recently, the relativistic approach was extended to describe the thermodynamics of color superconductivity by Blaschke et al.~\cite{blaschke}. At finite isospin density, by taking into account all the possible channels in the bubble summation in the RPA, the NJL model describes successfully the pion superfluid phase,  especially the Goldstone mode corresponding to the spontaneous isospin symmetry breaking~\cite{he1}.

In this paper, we study the quark-antiquark scattering phase shift and its relation to the meson spectral function in pion superfluid phase in the NJL model. When isospin symmetry is spontaneously broken, the new eigenmodes of the superfluid system are no longer the ordinary mesons with fixed isospin quantum numbers but their linear combinations. Such strong meson mixing makes it impossible to define quark-antiquark scattering phase shift and meson spectral function for a fixed channel, and the whole phase shift and spectral function are not a simple sum of all the possible meson channels. We will discuss the relation among the meson propagator pole, quark-antiquark scattering phase shift and meson spectral function in the NJL model in Section \ref{s2}, and show numerical calculations in normal and pion superfluid phases and analyze the physics behind in Section \ref{s3}. Finally we summarize in Section \ref{s4}.

\section{Meson Spectra in Pion Superfluid}
\label{s2}
The Lagrangian density of the two flavor NJL model at
quark level is defined as~\cite{klevansky}
\begin{equation}
\label{njl}
{\cal L} =
\bar{\psi}\left(i\gamma^{\mu}\partial_{\mu}-m_0+\mu\gamma_0\right)\psi
+G\Big[\left(\bar{\psi}\psi\right)^2+\left(\bar\psi
i\gamma_5{\vec \tau}\psi\right)^2\Big]
\end{equation}
with scalar and pseudoscalar interactions corresponding to
$\sigma$ and $\pi$ excitations, where $m_0$ is the current quark
mass, $G$ is the four-quark coupling constant with dimension
(GeV)$^{-2}$, $\tau_i\ (i=1,2,3)$ are the Pauli matrices in flavor
space, and $\mu
=diag\left(\mu_u,\mu_d\right)=diag\left(\mu_B/3+\mu_I/2,\mu_B/3-\mu_I/2\right)$
is the quark chemical potential matrix with $\mu_u$ and $\mu_d$
being the $u$- and $d$-quark chemical potentials and $\mu_B$ and
$\mu_I$ being the baryon and isospin chemical potentials.

At zero isospin chemical potential, the Lagrangian density has the
symmetry of\\
$U_B(1)\bigotimes SU_I(2)\bigotimes SU_A(2)$
corresponding respectively to baryon number symmetry, isospin symmetry and
chiral symmetry. At finite isospin chemical potential but without pion condensation, the isospin
symmetry $SU_I(2)$ and chiral symmetry $SU_A(2)$ are explicitly
broken to $U_I(1)$ and $U_A(1)$, respectively. Therefore, the
chiral symmetry restoration at finite isospin chemical potential
means only degeneracy of $\sigma$ and $\pi_0$ mesons, the charged
$\pi_+$ and $\pi_-$ behave still differently.

Introducing the chiral and pion condensates $\langle\sigma\rangle$ and $\langle\pi\rangle$ which are the order parameters to describe respectively the spontaneous chiral and isospin symmetry breaking,
\begin{eqnarray}
\label{condensate}
&& \langle\sigma\rangle = \langle\bar{\psi}\psi\rangle, \nonumber\\
&& \langle\pi\rangle =\sqrt 2\langle\bar{\psi}i\gamma_5\tau_\pm\psi\rangle
\end{eqnarray}
with $\tau_\pm =(\tau_1\pm i\tau_2)/\sqrt 2$, and taking both
condensates to be real, the quark propagator in mean field
approximation can be expressed as a matrix in flavor space~\cite{he1},
\begin{eqnarray}
\label{quarkpropagator1}
{\cal S}^{-1}\left(k\right)=\left(\begin{matrix}\gamma^\mu k_\mu+\mu_u\gamma_0-M_q & 2iG\langle\pi\rangle\gamma_5\cr 2iG\langle\pi\rangle\gamma_5 & \gamma^\mu k_\mu+\mu_d\gamma_0-M_q\cr
\end{matrix}\right)
\end{eqnarray}
with off diagonal elements in the phase of spontaneous isospin symmetry breaking. The chiral symmetry breaking is hidden in the effective quark mass
$M_q=m_0-2G\langle\sigma\rangle$. Using the method of massive energy projector~\cite{huang}, one can explicitly obtain the mean field quark propagator~\cite{he1}
\begin{eqnarray}
\label{quarkpropagator2}
{\cal S}\left(k\right)=\left(\begin{matrix}{\cal S}_{uu}(k) & {\cal S}_{ud}(k)\cr {\cal S}_{du}(k) & {\cal S}_{dd}(k)\cr
\end{matrix}\right)
\end{eqnarray}
in terms of the four effective quark energies
\begin{equation}
E^\pm_\mp = \sqrt{\left(\sqrt{{\bf k}^2+M_q^2}\pm {\mu_I\over 2}\right)^2+4G^2\langle\pi\rangle^2}\mp{\mu_B\over 3}.
\label{energy}
\end{equation}
From the definitions of the chiral and pion condensates (\ref{condensate}), it is easy to express them in terms of the matrix elements of the quark propagator,
\begin{eqnarray}
\label{gap}
\langle\sigma\rangle &=&  -N_c\int{d^4k\over (2\pi)^4}\text{Tr}_D\left[i{\cal S}_{uu}(k)+i{\cal S}_{dd}(k)\right],\nonumber\\
\langle\pi\rangle &=&  N_c\int{d^4k\over (2\pi)^4}\text{Tr}_D\left[\left({\cal S}_{ud}(k)+{\cal S}_{du}(k)\right)\gamma_5\right],
\end{eqnarray}
where the trace $\text {Tr}_D$ is taken in Dirac space, and the
four momentum integration is defined as $\int d^4k/(2\pi)^4=iT\sum_n\int d^3{\bf k}/(2\pi)^3$ in Euclidean space with $k_0=i\omega_n=i(2n+1)\pi T\ (n=0,\pm 1,\pm 2,\cdots)$ at finite temperature $T$. Obviously, the
color degrees of freedom in the NJL model is trivial, and
the trace in color space simply contributes a color factor $N_c=3$. The two gap equations (\ref{gap}) determine self-consistently the two order parameters $\langle\sigma\rangle(T,\mu)$ and $\langle\pi\rangle(T,\mu)$ in medium. In chiral limit with vanishing current quark mass $m_0=0$, the critical point of chiral symmetry restoration is determined by the first gap equation (\ref{gap}) at $\sigma=0$, and the pion superfluid starts at $\mu_I=0$.
In real case with $m_0\neq 0$, while there is no more strict chiral phase transition, the pion superfluid starts at $\mu_I=M_\pi$ at $T=0$~\cite{he1}, where $M_\pi$ is the pion mass in vacuum.

The above treatment at mean field level is clearly incomplete, it misses the meson contribution which should dominate the system at low temperature and chemical potential. In the NJL model, the meson modes are regarded as quantum
fluctuations above the mean field. The quark-antiquark scattering via
meson exchange can be effectively expressed at quark level in
terms of quark bubble summation in RPA~\cite{hufner,zhuang}
. Considering all the possible quark bubbles between two interaction vertexes, the meson propagator can then be written as a matrix in the meson space $(\pi_+,\pi_-,\pi_0,\sigma)$~\cite{he1,xia},
\begin{equation}
D(p)={2G\over 1-2G\Pi(p)},
\end{equation}
where $p=(i\nu_n,{\bf p})$ is the meson four momentum with the Matsubara frequency $\nu_n=2\pi
nT$, and $\Pi(p)$ is the meson polarization matrix with elements
\begin{equation}
\label{element}
\Pi_{mn}(p)=i\int {d^4q\over (2\pi)^4}\text{Tr}\left[\Gamma_m {\cal S}(p+q)\Gamma_n^*{\cal S}(q)\right]
\end{equation}
with the meson vertexes
\begin{equation}
\label{vertex} \Gamma_m = \left\{\begin{array}{ll}
1 & m=\sigma\\
i\tau_+\gamma_5 & m=\pi_+ \\
i\tau_-\gamma_5 & m=\pi_- \\
i\tau_3\gamma_5 & m=\pi_0\ ,
\end{array}\right.\ \
\Gamma_m^* = \left\{\begin{array}{ll}
1 & m=\sigma\\
i\tau_-\gamma_5 & m=\pi_+ \\
i\tau_+\gamma_5 & m=\pi_- \\
i\tau_3\gamma_5 & m=\pi_0\ . \\
\end{array}\right.
\end{equation}
In Eq.(\ref{element}) the quark momentum integration is at finite temperature, and the trace is over color, flavor and spin degrees of freedom of quarks. It is easy to see that the polarization matrix is symmetric with $\Pi_{mn}=\Pi_{nm}$. In normal phase with vanishing pion condensate $\langle\pi\rangle=0$, the quark propagator is diagonal in flavor space, and therefore all the off-diagonal elements of the polarization matrix $\Pi$ in the meson space disappear automatically. In the pion superfluid, while there
is no mixing between $\pi_0$ and other mesons, $\Pi_{\pi_0\sigma} = \Pi_{\pi_0\pi_+} = \Pi_{\pi_0\pi_-} =
0$, the other three mesons $\sigma, \pi_+$ and $\pi_-$ are coupled to each other with nonzero elements $\Pi_{\sigma\pi_+}, \Pi_{\sigma\pi_-}, \Pi_{\pi_+,\pi_-}\neq 0$~\cite{he1}.

The static meson properties like mass and the meson contribution to the thermodynamics of the system~\cite{hufner,zhuang} are controlled by the determinant of the inverse meson propagator which is a complex function and can be expressed in terms of a phase,
\begin{eqnarray}
\label{phase1}
&& \text {det}\left[1-2G\Pi(\omega\pm i\epsilon,{\bf p})\right] = \left|\text {det}\left[1-2G\Pi(\omega + i\epsilon,{\bf p})\right]\right|e^{\mp i\Phi},\nonumber\\
&& \tan \Phi = -{\text{Im}\{\text {det}\left[1-2G\Pi(\omega + i\epsilon,{\bf p})\right]\}\over \text {det}\left[1-2G\Pi(\omega,{\bf p})\right]}.
\end{eqnarray}
Physically, $\Phi$ is the phase shift associated with the quark-antiquark scattering in the model. While the phase shift is controlled by the meson mass $M$ around a pole of the propagator which is determined by $\text{det}\left[1-2G\Pi(M,{\bf p})\right]=0$, it is dominated by the background when the kinematics is away from the pole. In normal phase without pion condensate, it can be explicitly separated into a meson part and a background part, and the latter is independent of the meson itself and controlled by the quark properties~\cite{zhuang}.

The meson spectral function which is also controlled by the complex structure of the meson propagator is closely related to the quark-antiquark scattering phase shift. In the case without meson mixing, the spectral function for the meson $m$ in the NJL model is conventionally defined as
\begin{eqnarray}
\label{spectra1}
\rho_m(\omega,{\bf p}) &=& -2\text {Im} D_{mm}(\omega+i\epsilon,{\bf p})\\
&=& -{2G\sin \left(2\Phi_m(\omega,{\bf p})\right)\over 1-2G\Pi_{mm}(\omega,{\bf p})}\nonumber
\end{eqnarray}
with the scattering phase shift for the meson $m$,
\begin{equation}
\label{phase2}
\tan \Phi_m = -{\text{Im}\left[1-2G\Pi_{mm}(\omega + i\epsilon,{\bf p})\right]\over 1-2G\Pi_{mm}(\omega,{\bf p})}.
\end{equation}
In the pion superfluid phase with $\pi_+-\pi_--\sigma$ mixing, one can not separately define the spectral function for a fixed channel, the whole spectral function of the mixed mesons is defined through the whole mixed meson propagator,
\begin{eqnarray}
\label{spectra2}
\rho(\omega,{\bf p}) &=&-2\text {Im}\{\text {det}D(\omega+i\epsilon,{\bf p})\}\\
&=&-2\text {Im}\left[{\text {det}(2G)\over \text {det}\left[1-2G\Pi(\omega+i\epsilon,{\bf p})\right]}\right]\nonumber\\
&=&-{(2G)^3\sin \left(2\Phi(\omega,{\bf p})\right)\over \text {det} \left[1-2G\Pi(\omega,{\bf p})\right]}\nonumber.
\end{eqnarray}
In any case, the meson spectral function is associated with both the meson pole shown in the denominator and the quark-antiquark scattering phase shift shown in the numerator. This means that in the quark-meson plasma described by the NJL model, the meson spectral function is governed by not only the meson characteristics but also the quark properties.

\section{Numerical Results and Discussions}
\label{s3}
Before doing numerical calculations, we first fix the parameters of the model.
Since there is no confinement in the NJL model and the model is non-renormalizable, we should
regularize the model. For simplicity, we choose a hard cutoff $\Lambda$ for the quark three momentum. The three parameters $\Lambda$, $G$ and $m_0$
are fixed by fitting the physical quantities in vacuum: pion mass $M_\pi=0.134$ GeV,
pion decay constant $F_\pi=0.093$ GeV and the chiral condensate
$\langle\sigma\rangle=2(-0.25\text {GeV})^3$. The parameters we obtained are $\Lambda=0.653$ GeV,
$G=4.93$ GeV$^{-2}$ and $m_0=0.005$ GeV~\cite{zhuang}.

For mesons in stable bound states, their masses $M_m$ are determined by the poles of the meson propagator at vanishing three momentum,
\begin{equation}
\label{pole1}
\text {det}\left[1-2G\Pi(M_m,{\bf 0})\right]=0.
\end{equation}
Around the pole $\omega^2=E_m^2=M_m^2+{\bf p}^2$, the imaginary part of the determinant can be neglected, and the scattering phase shift is reduced to a simple step function,
\begin{equation}
\label{step1}
\Phi_m(\omega,{\bf p})=\pi\Theta\left(\omega^2-E_m^2\right),
\end{equation}
and its contribution to the thermodynamic potential becomes the familiar one for an ideal meson gas,
\begin{equation}
\label{bound}
\Omega_m=\int\frac{d^3 {\bf p}}{(2\pi)^3}\left[\frac{E_m}
    {2}+\frac{1}{\beta}\text{ln}\left(1-e^{-\beta E_m}\right)\right]
\end{equation}
with $\beta=1/T$.

However, when the meson mass exceeds two times the quark mass, the meson can decay into its constituent quark-antiquark pair. It is thus not a stable bound state, but rather a resonant state. In this case, the pole is no longer on the real axis of the $\omega$-plane, the pole equation (\ref{pole1}) should be regarded in its complex form in order to determine the resonant mass $M_m$ and its associated width $\Gamma_m$ that are defined through the relation
\begin{equation}
\label{pole2}
\text {det}\left[1-2G\Pi(M_m-i\Gamma_m/2,{\bf 0})\right]=0.
\end{equation}
Around the complex pole $\omega^2=\left(M_m-i\Gamma_m/2\right)^2+{\bf p}^2$, the imaginary part of the determinant can be considered as a constant and the scattering phase shift satisfies
\begin{equation}
\label{step2}
\tan \Phi_m(\omega,{\bf p})=-{M_m\Gamma_m\over \omega^2-\left(E_m^2-\Gamma_m^2/4\right)}.
\end{equation}
\begin{figure}[htb]
\begin{center}
\includegraphics[scale=0.8]{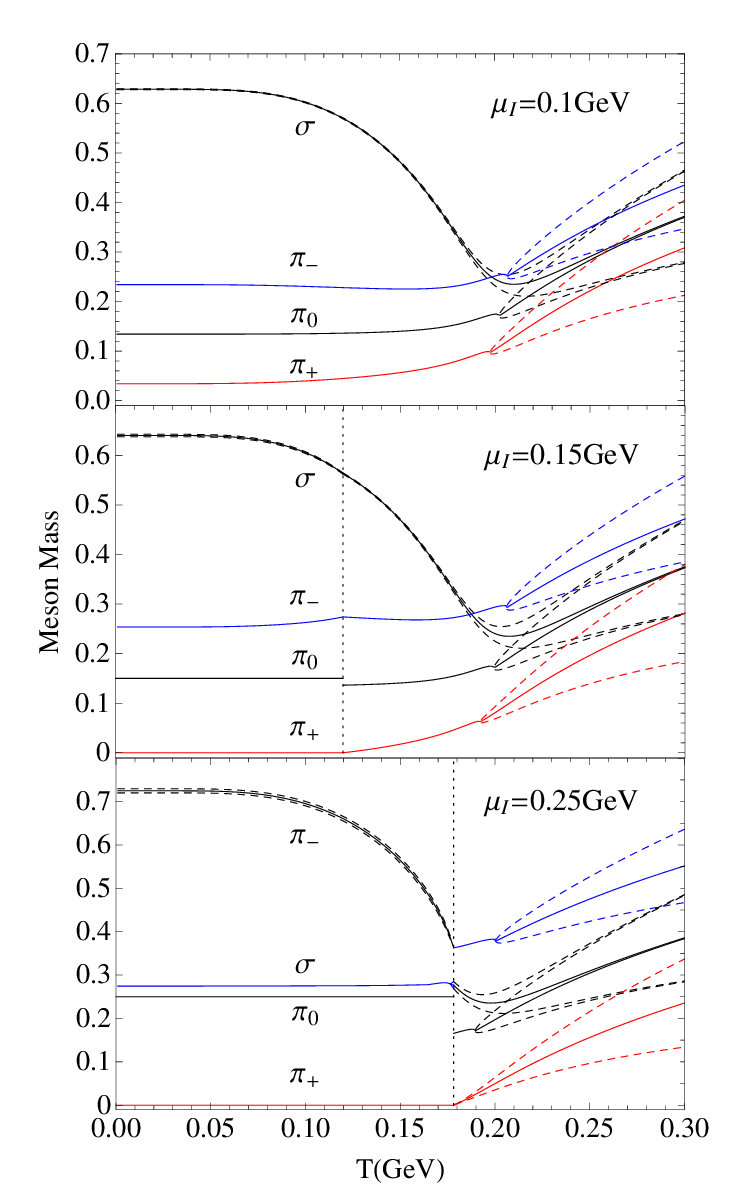}
\end{center}
\caption{ (color online) The meson masses $M_m(T)$ (solid lines) and their spreadings $M_m(T)\pm\Gamma_m(T)/2$ (dashed lines) at different isospin chemical potential $\mu_I=$0.1 (upper panel), 0.15 (middle panel), 0.25 (lower panel) GeV and vanishing baryon chemical potential $\mu_B=0$. The vertical dotted lines in the middle and lower panels separate the pion superfluid phase at low temperature from the normal phase at high temperature. }
\label{fig1}
\end{figure}

The meson masses $M_m$ and the associated widthes $\Gamma_m$ are displayed in Fig.\ref{fig1} as functions of temperature at fixed isospin chemical potential. At $\mu_I=0.1$ GeV (upper panel) which is less than the critical value $\mu_I=M_\pi$ of the pion superfluid, the system is in normal phase without pion condensate at any temperature. Due to the explicit isospin symmetry breaking from $SU_I(2)$ to $U_I(1)$ at $\mu_I\neq 0$, the degenerated pion mass $M_\pi$ in vacuum splits into $M_{\pi_+},\ M_{\pi_-}$ and $M_{\pi_0}$ in medium. On the other hand, the explicit chiral symmetry breaking from $SU_A(2)$ to $U_A(1)$ at $\mu_I\neq 0$ reduces the degree of meson degeneration at high temperature: only the neutral mesons $\sigma$ and $\pi_0$ become degenerate, the charged pions $\pi_+$ and $\pi_-$ behave still differently. Any pion is in its bound state at low temperature (especially in the pion superfluid phase) and starts to have nonzero width at the corresponding critical temperature of Mott phase transition~\cite{mott,hufner2,costa} where the pion energy is larger than the corresponding quark plus antiquark energies. The width increases monotonously with temperature. The $\sigma$ meson is always in the resonant state in the NJL model, the width is however rather small in the chiral symmetry breaking phase and increases rapidly when the symmetry is gradually restored. Different from the case at $\mu_I=0$ where the three pions are degenerate, the meson mass splitting at $\mu_I\neq 0$ leads to different Mott phase transition points for different pions.

In the normal phase with only chiral condensation, the quark propagator is diagonal and the summation of bubbles in RPA selects its specific channel by choosing
at each stage the same proper polarization function. Therefore, a
meson mode which is determined by the pole of the corresponding
meson propagator is related to its own polarization
function only. When however the isospin chemical potential exceeds the critical value $\mu_I=M_\pi$, see the middle and lower panels of Fig.\ref{fig1}, the system enters the pion superfluid phase at low temperature. The collective
excitations of the system, determined by the pole equation (\ref{pole2}), will in principle not be the
original meson modes ($\pi_+,\pi_-,\pi_0,\sigma$) but their linear combination ($c_{\pi_+}\pi_++c_{\pi_-}\pi_-+c_{\pi_0}\pi_0+c_\sigma\sigma$)~\cite{he1}. Due to this mixing which is very strong at high $\mu_I$ and low $T$, these new modes in the pion superfluid are no longer the eigenmodes of the isospin operator $\hat I_3$. In the NJL model, from the explicit calculation of the polarization elements $\Pi_{mn}$, $\pi_0$ decouples from the other mesons and is still an collective mode of the system, but the other three ($\pi_+,\pi_-,\sigma$) are replaced by their linear combinations. Considering the fact that these new modes should come back to the old modes at the critical point of pion superfluid, indicated by the vertical dotted lines in the middle and lower panels of Fig.\ref{fig1}, we still name these modes $\pi_+,\pi_-,\sigma$ according to the continuity. While the mixing is weak around the critical point, it becomes stronger and stronger at $T\to 0$. For instance, at $\mu_I=0.25$ GeV, see the lower panel of Fig.\ref{fig1}, the new mode $\pi_-$ in the pion superfluid is dominated by the $\pi_-$ component at $T\to T_c$ but by the $\sigma$ component at $T\to 0$. Corresponding to the spontaneous isospin symmetry breaking, there exists a Goldstone mode in the pion superfluid. From the continuity at the critical point, we call it $\pi_+$. While there is a small width associated with the new mode $\sigma$ in the middle panel and $\pi_-$ in the lower panel, the other new modes are in the bound states with $\Gamma = 0$. The critical temperature starts with $T=0$ at $\mu_I=M_\pi$ and increases with $\mu_I$ monotonously. In the normal phase at high temperature, again any meson becomes a resonant state when its energy is beyond the quark plus antiquark energies. The meson widthes increase rapidly with temperature.

For $\pi_0$ meson which decouples from the other three mesons, we can analytically derive
\begin{equation}
\label{pi0}
M_{\pi_0}(T,\mu_I)=\mu_I
\end{equation}
in the pion superfluid, by combining the pole equation $1-2G\Pi_{\pi_0\pi_0}(M_{\pi_0},{\bf 0})=0$ and the gap equation for the pion condensate $\langle\pi\rangle$. However, the polarization $\Pi_{\pi_0\pi_0}$ is not continuous at the critical point of pion superfluid in the case of nonzero temperature, leading to a jump of $M_{\pi_0}$ from $\mu_I$ in the pion superfluid phase to a lower value in the normal phase, see the jumps alone the vertical dotted lines in the middle and lower panels of Fig.\ref{fig1}. When temperature is high enough, the chiral symmetry is gradually restored, and the $\sigma$ and $\pi_0$ masses approach to coincide.
\begin{figure}[htb]
\begin{center}
\includegraphics[scale=1]{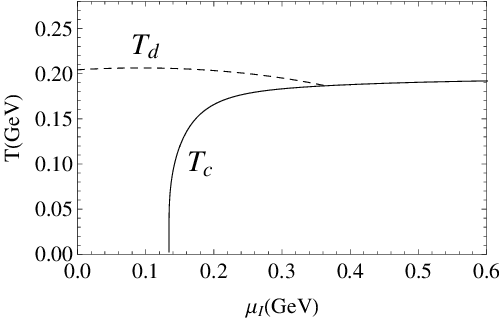}
\end{center}
\caption{The phase diagram of pion superfluid in the $T-\mu_I$ plane at $\mu_B=0$. The solid and dashed lines display respectively the critical temperature $T_c$ of the pion superfluid and the dissociation temperature $T_d$ of the meson bound states.}
\label{fig2}
\end{figure}

With the gap equations (\ref{gap}) for the chiral and pion condensates $\langle\sigma\rangle$ and $\langle\pi\rangle$, one can determine the phase boundaries of chiral restoration and pion superfluid in the NJL model, and with the pole equation (\ref{pole1}) in the normal phase, one can obtain the border between the meson bound and resonant states. Fig.\ref{fig2} shows the phase diagram of the pion superfluid in the plane of temperature and isospin chemical potential at vanishing baryon chemical potential. The critical temperature $T_c$ and the $\pi_0$ dissociation temperature $T_d$ are displayed by solid and dashed lines, respectively. The two temperatures coincide when $\mu_I$ is large enough. There are three phases in the plane: the pion superfluid phase in the region of $T<T_c$, the normal phase with meson bound states in the region of $T_c<T<T_d$, and the normal phase without bound states in the region of $T>T_d$. Since the thermal excitation of the pion condensate at the critical temperature $T_c$ are meson bound states at low $\mu_I$ and resonant states or even quark and antiquark states at high $\mu_I$, there should exist in the pion superfluid phase a crossover from Bardeen-Cooper-Schrieffer (BCS) state at high $\mu_I$ to Bose-Einstein Condensation (BEC) state at low $\mu_I$~\cite{he2}.
\begin{figure}[htb]
\begin{center}
\includegraphics[scale=1]{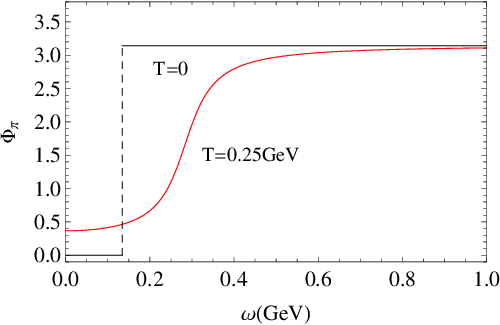}
\end{center}
\caption{ (color online) The pion phase shift $\Phi_\pi$ as a function of pion energy $\omega$ at vanishing chemical potentials and momentum $\mu_I=\mu_B={\bf p}=0$. The two lines correspond to the bound state at $T=0$ and resonant state at $T=0.25$ GeV. }
\label{fig3}
\end{figure}

We now consider the quark-antiquark scattering phase shift in pole approximation. For all the numerical calculations for phase shifts and spectral functions shown in the following, we set the meson momentum ${\bf p}=0$. At $\mu_I=0$, pions are degenerate and in bound state at low temperature and resonant state at high temperature. The pion phase shift $\Phi_\pi(\omega)$ in pole approximation is shown in Fig.\ref{fig3} for the bound and resonant states. In vacuum at $T=0$, the phase shift $\Phi_\pi (\omega)$ is a step function. It is zero at $\omega < M_\pi$, jumps suddenly from $0$ to $\pi$ at $\omega=M_\pi$, and keeps $\pi$ at $\omega > M_\pi$. At $T=0.25$ GeV, the pions are already in resonant state, and the phase shift changes from the step function (\ref{step1}) to the smooth crossover (\ref{step2}). The rapid change is still around $\omega=M_\pi= 0.28$ GeV. While it can reach the limit $\Phi_\pi=\pi$ at large enough $\omega$, it starts with a finite value $\Phi_\pi\neq 0$ at $\omega=0$ due to the finite width $\Gamma_\pi$.
\begin{figure}[htb]
\begin{center}
\includegraphics[scale=0.9]{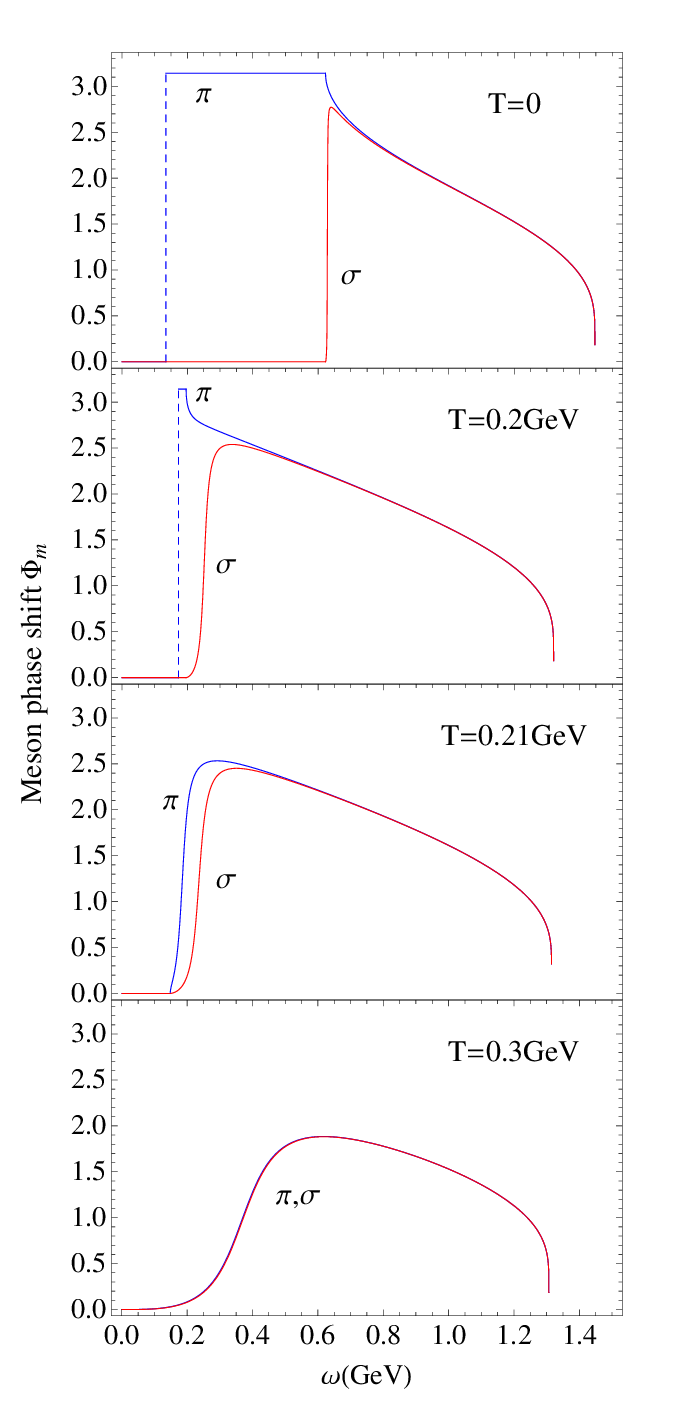}
\end{center}
\caption{ (color online) The full scattering phase shifts $\Phi_m(\omega)\ (m=\pi, \sigma)$ in normal phase at $T=0,\ 0.2,\ 0.21,\ 0.3$ GeV and $\mu_B=\mu_I=0$.}
\label{fig4}
\end{figure}

However, the pole approximation is in principle insufficient for our calculation, since it is only valid for $\omega$ in the vicinity of the pole and should violently deviate from the original definition (\ref{phase1}). Furthermore, it neglects the fact, that the exact phase shift $\Phi(\omega)$ must increase around the pole and then decrease in such a way, that Levinson's theorem be fulfilled~\cite{hufner}. In our notation this theorem reads
\begin{equation}
\label{levinson}
\int_0^\infty d\omega{d\Phi\over d\omega}=0.
\end{equation}
Thus the pole approximation, from which the phase shift starts with $\Phi=0$ and ends with $\Phi=\pi$, is inconsistent with Levinson's theorem, and we therefore expect that a partial compensation to the meson spectral function $\rho$ as calculated from (\ref{spectra2}) must arise when one performs a full calculation of the phase shift. Fig.\ref{fig4} shows the full scattering phase shift in normal phase at temperature $T=0,\ 0.2,\ 0.21$ and $0.3$ GeV and vanishing chemical potentials $\mu_B=\mu_I=0$. The full phase shift includes not only the meson part which is controlled by the step function (\ref{step1}) or the crossover (\ref{step2}) around $\omega=M_m$ but also a background part which starts at the threshold $\omega_{th} = \text {Min}(E^-_-+E^-_+)=2M_q$ and is independent of the meson properties, where $E^-_\pm$ are the quasi particle energies defined in (\ref{energy}). In the normal phase at $\mu_I=0$ this has been explicitly proven~\cite{zhuang}. It is the background part which makes the full phase shift satisfy the levinson's theorem (\ref{levinson}): the phase shift starts with $\Phi(\omega=0)=0$ and ends with $\Phi(\omega\to\infty)=0$.
The rapid decrease of the phase shift at $\omega_{max}=\text {Max}(E^-_-+E^-_+)=2\sqrt{M_q^2+\Lambda^2}$ is associated to the finite momentum cutoff $\Lambda$ in the model. The phase shift retains zero for $\omega>\omega_{max}$. In vacuum with $T=\mu_B=\mu_I=0$, the degenerated pions are in bound state and the phase shift is given by the step function around the pion mass and then drops down continuously due to the contribution from the background. For sigma, its mass is slightly larger than two times the quark mass in the model, it is nearly a bound state. However, before the phase shift $\Phi_\sigma$ reaches $\pi$ at $\omega=M_\sigma$, the decreasing background part starts already at $\omega=2M_q<M_\sigma$. Therefore, the full phase shift can not reach $\pi$. Since the background contribution is meson independent, the phase shifts for $\pi$ and $\sigma$ coincide at large $\omega$.
\begin{figure}[htb]
\begin{center}
\includegraphics[scale=0.9]{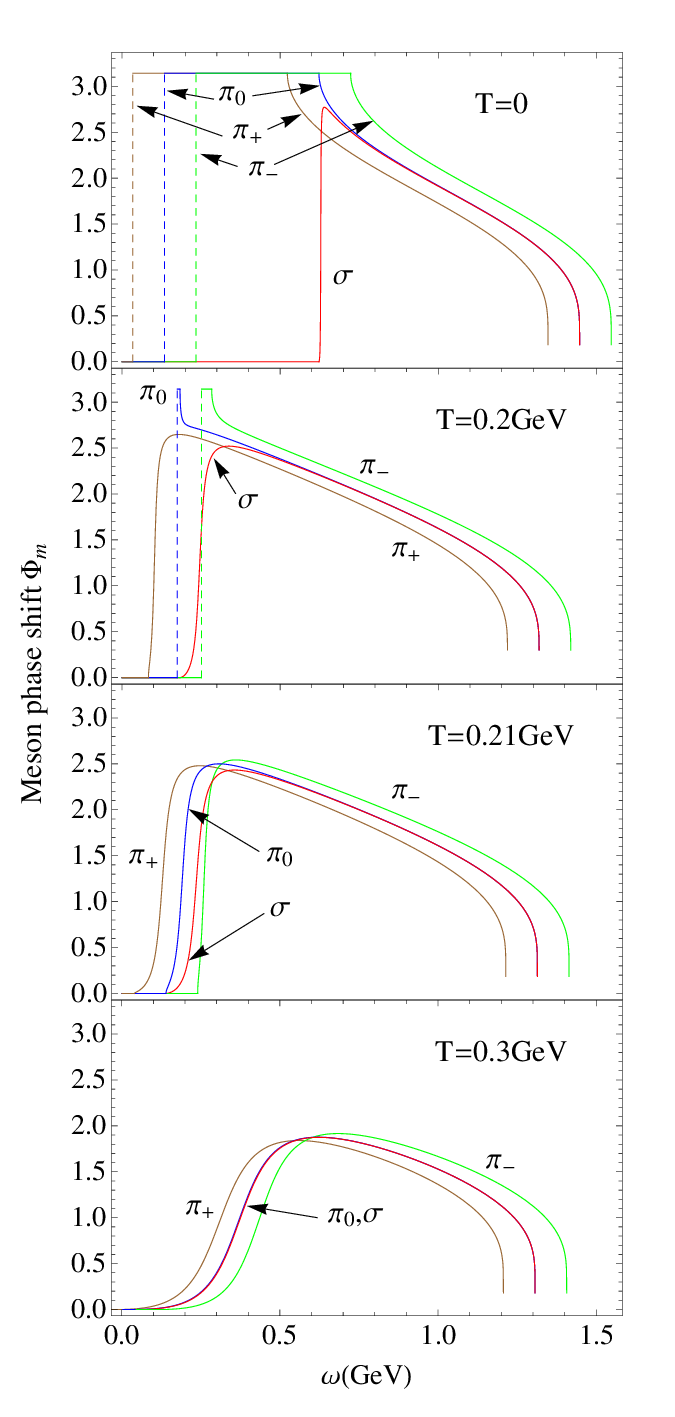}
\end{center}
\caption{ (color online) The full scattering phase shifts $\Phi_m(\omega)\ (m=\pi_+, \pi_-, \pi_0, \sigma)$ in normal phase at $T=0,\ 0.2,\ 0.21,\ 0.3$ GeV, $\mu_B=0$ and $\mu_I=0.1$ GeV. }
\label{fig5}
\end{figure}

At $T=0.2$ GeV which is below the critical temperature for the Mott phase transition of pions, the behavior of the phase shifts for pions and sigma is similar to that at $T=0$. The pions are still in bound state with larger mass, their phase shift reaches $\pi$ at $\omega=M_\pi$ and then decreases continuously and finally approaches zero at $\omega_{max}$, due to the contribution from the background. The sigma becomes much lighter but still in a resonant state with $M_\sigma>2M_q$. At $T=0.21$ GeV which is already larger than the critical temperature for the Mott phase transition of pions, both pions and sigma are in resonant states with $M_\pi,\ M_\sigma >2M_q$, and both phase shifts can not reach $\pi$. At $T=0.3$ GeV, the chiral symmetry is well restored with a small quark mass, and all the mesons have almost the same mass. In this case the phase shifts for pions and sigma coincide in the whole $\omega$ region. Since the contribution from the background starts very early, the strong cancelation between the increasing meson part and the decreasing background part leads to a rather small phase shift.

The full phase shifts at $\mu_I=0.1$ GeV are shown in Fig.\ref{fig5}. Since $\mu_I$ is less than the critical isospin chemical potential $\mu_I=M_\pi$, the isospin symmetry is explicitly broken but the system is still in normal phase without pion condensate. The pions at low temperature are still in bound states but with mass splitting. In vacuum with $T=0$, the mass splitting is $\Delta M_\pi=M_{\pi_0}-M_{\pi_+}=M_{\pi_-}-M_{\pi_0}=\mu_I$. The continuous background contribution to $\pi_+$ starts at $\omega_{th}=\text {Min}(E^-_-+E^-_+)=2M_q-\mu_I$ and ends at $\omega_{max}=\text {Max}(E^-_-+E^-_+)=2\sqrt{M_q^2+\Lambda^2}-\mu_I$. Due to the meson independence of the background contribution, the phase shifts for the isospin neutral mesons $\pi_0$ and $\sigma$ coincide at large $\omega$. After a shift of $\omega\to\omega\pm\mu_I$ for $\pi_+$ and $\pi_-$, all the phase shifts coincide at large $\omega$. With increasing temperature, the $\pi_+$ meson first becomes a resonant state at $T=0.2$ GeV and then immediately the other two pions $\pi_0$ and $\pi_-$ get widthes at $T=0.21$ GeV. When the temperature is high enough, for instance $T=0.3$ GeV, all the scattering phase shifts coincide after proper shifts.

\begin{figure}[htb]
\begin{center}
\includegraphics[scale=0.9]{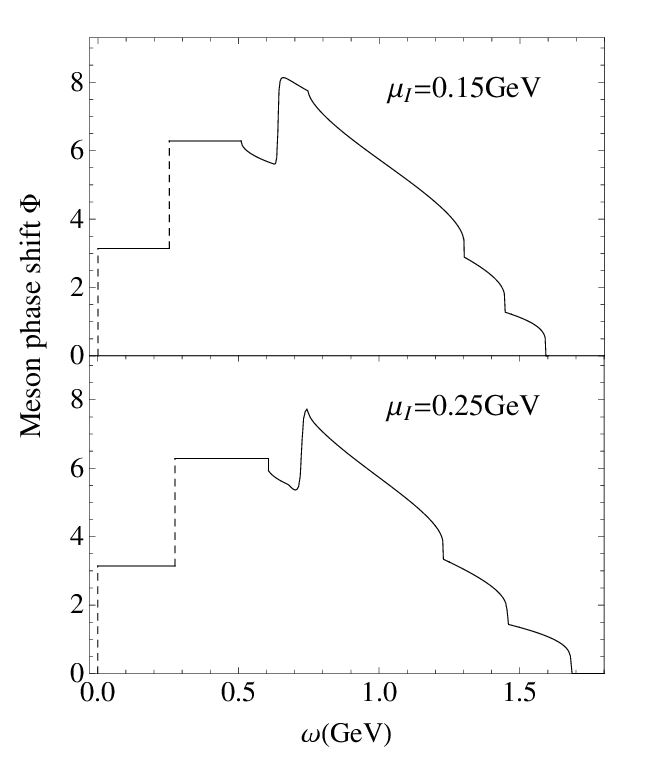}
\end{center}
\caption{The whole quark-antiquark scattering phase shift $\Phi(\omega)$ in pion superfluid phase at $\mu_I=0.15,\ 0.25$ GeV and $T=\mu_B=0$. }
\label{fig6}
\end{figure}

In the pion superfluid phase all the four mesons are in bound or nearly bound states, see Fig.\ref{fig1}. Since the meson $\pi_0$ decouples from the other three mesons, its phase shift is similar to the one in normal phase shown in Fig.\ref{fig5}, the only change is the location of the jump and the threshold for the continuous background contribution. Because of the mixing among $\pi_+, \pi_-$ and $\sigma$, one can not separately define their independent phase shifts. The whole phase shift for the three mixed mesons in pion superfluid is shown in Fig.\ref{fig6} at finite isospin chemical potential and vanishing temperature and baryon chemical potential. While it looks like a sum of the three independent phase shifts for the three new eigenmodes (we still call them $\pi_+, \pi_-$ and $\sigma$) of the system, the pion condensate changes dramatically the contribution from the continuous background. The first jump from $0$ to $\pi$ happens at $\omega=M_{\pi_+}=0$ (note that $\pi_+$ is the Goldstone mode), and the second jump happens at $\omega=M_{\pi_-}=0.25$ GeV for $\mu_I=0.15$ GeV (the upper panel) and $\omega=M_\sigma=0.27$ GeV for $\mu_I=0.25$ GeV (the lower panel), corresponding to the meson masses shown in Fig.\ref{fig1}. Before the third jump for the mode $\sigma\ (\pi_-)$ at $\omega=0.64\ (0.72)$ GeV, the continuous contribution from the background starts already which cancels partly the third jump. The background contribution starts at
\begin{eqnarray}
\label{threshold}
\omega_{th} &=& \text {Min}(E^-_-+E^-_+)\nonumber\\
&=& \text {Min}\left(2\sqrt{\left(\sqrt{{\bf k}^2+M_q^2}-{\mu_I\over 2}\right)^2+4G^2\langle\pi\rangle^2}\right).
\end{eqnarray}
For $M_q>\mu_I/2$, like the upper panel of Fig.\ref{fig6}, the minimum is at vanishing quark momentum ${\bf k}=0$, and the threshold is $\omega_{th}=2\sqrt{\left(M_q-\mu_I/2\right)^2+4G^2\langle\pi\rangle^2}$. For $M_q<\mu_I/2$, like the lower panel of Fig.\ref{fig6}, the threshold is $\omega_{th}=4G\langle\pi\rangle$ but with a finite quark momentum $|{\bf k}|=\sqrt{\mu_I^2/4-M_q^2}$. Due to the momentum cutoff in the model, the full phase shift drops down sequentially at the maximum values
\begin{eqnarray}
\label{maximum}
\omega_{max} &=& \text {Max}(E^\pm_- + E^\pm_+)\nonumber\\
&=& \left\{\begin{array}{ll}
2\sqrt{\left(\sqrt{\Lambda^2+M_q^2}-{\mu_I\over 2}\right)^2+4G^2\langle\pi\rangle^2} & \\
2\sqrt{\Lambda^2+M_q^2+4G^2\langle\pi\rangle^2} & \\
2\sqrt{\left(\sqrt{\Lambda^2+M_q^2}+{\mu_I\over 2}\right)^2+4G^2\langle\pi\rangle^2} & ,
\end{array}\right.
\end{eqnarray}
which correspond to the three sudden drops of the phase shift at large $\omega$ shown in Fig.\ref{fig6}.
\begin{figure}[htb]
\begin{center}
\includegraphics[scale=0.9]{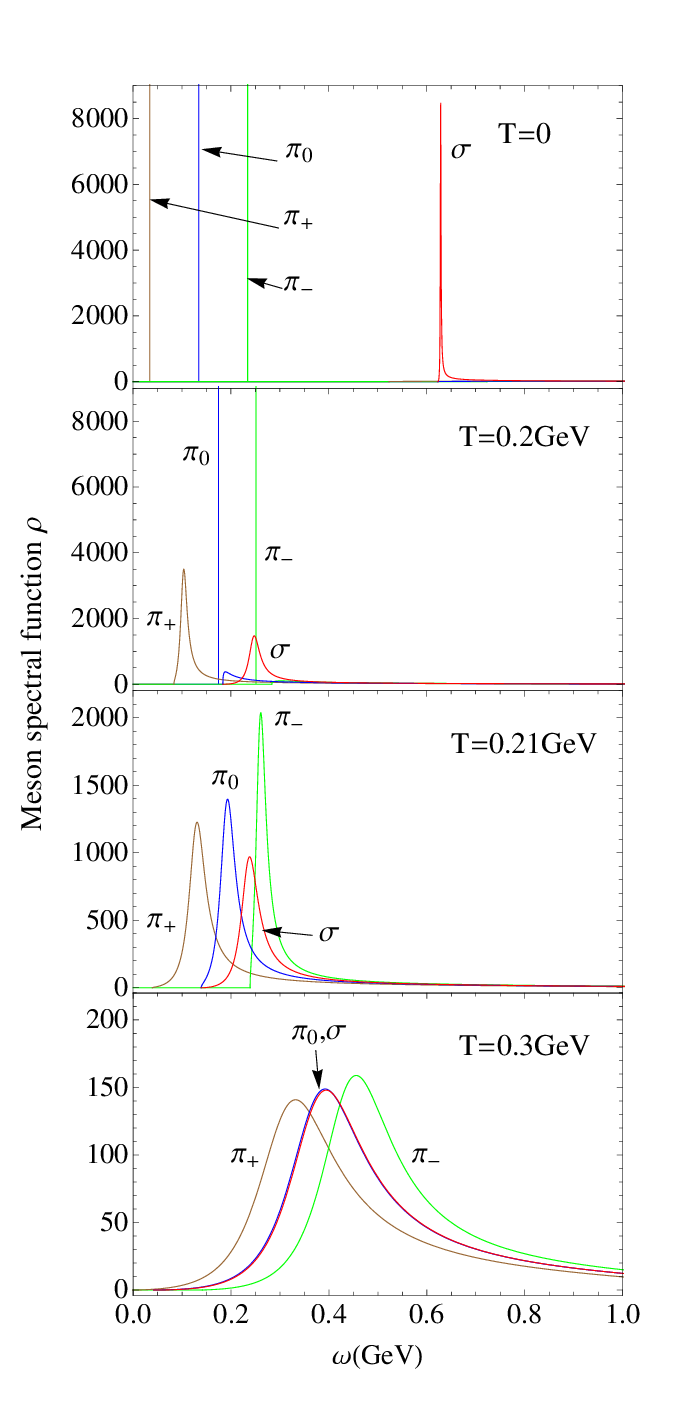}
\end{center}
\caption{ (color online) The meson spectral functions $\rho_m(\omega)\ (m=\pi_+, \pi_-, \pi_0, \sigma)$ in normal phase at $T=0,\ 0.2,\ 0.21,\ 0.3$ GeV, $\mu_B=0$ and $\mu_I=0.1$ GeV. }
\label{fig7}
\end{figure}
\begin{figure}[htb]
\begin{center}
\includegraphics[scale=1]{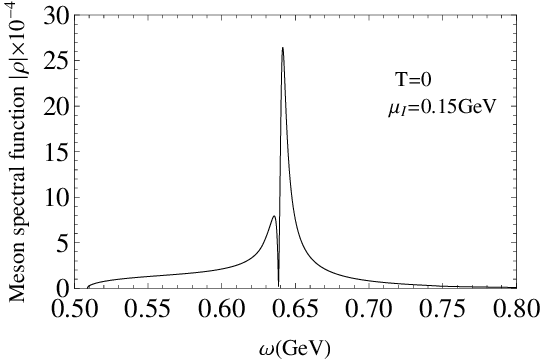}
\end{center}
\caption{The meson spectral function $\rho(\omega)$ in the pion superfluid phase at $\mu_I=0.15$ GeV and $T=\mu_B=0$. Only the part around the pole of the new mode $\sigma$ is displayed, and the two $\delta$ distributions at $\omega=0$ for the Goldstone mode $\pi_+$ and $0.25$ GeV for $\pi_-$ are not shown in the figure. }
\label{fig8}
\end{figure}
\begin{figure}[htb]
\begin{center}
\includegraphics[scale=1]{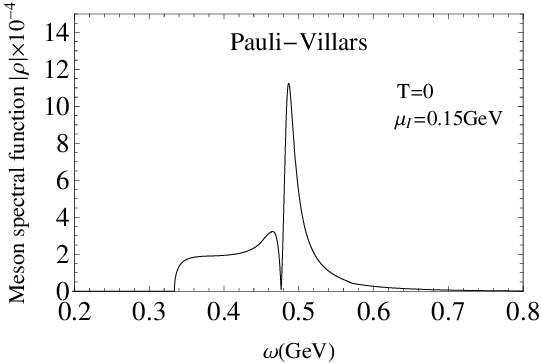}
\end{center}
\caption{The meson spectral function $\rho(\omega)$ in the pion superfluid phase at $\mu_I=0.15$ GeV and $T=\mu_B=0$, where only the part around the pole of the new mode $\sigma$ is displayed as in Fig.\ref{fig8}. Here, we use the Pauli-Villars regularization scheme in the calculation, and the new mode $\sigma$ has the mass $m_\sigma=0.48$ GeV. }
\label{fig9}
\end{figure}

The meson spectral function $\rho_m$ in normal phase is shown in Fig.\ref{fig7}. In principle, any meson spectral function should contain a continuous part, due to the phase shift from the background. However, when the meson is in bound state, the contribution from the scattering phase shift is strongly suppressed by the meson pole, see the cancelation between the phase shift in the numerator and the pole in the denominator of Eq. (\ref{spectra1}). At $T=0$, all the three pions are in bound states and their spectral functions are almost $\delta$ functions located at the corresponding poles, and the contribution from the background phase shift is very small and can be neglected in comparison with the pole. Since $\sigma$ is always in resonant state, its spectral function is already with a very small width in vacuum. However, the continuous phase shift which is mainly on the right hand side of the pole makes the spectral function no longer symmetric. With increasing temperature, the three pions will sequentially become in resonant states and all the meson widths increase continuously. When chiral symmetry is well restored, the $\pi_0$ and $\sigma$ spectral functions coincide. Note that, the asymmetry of the spectral function becomes more and more clear with increasing temperature.

In the pion superfluid phase, all the new meson modes $\pi_+, \pi_-$ and $\sigma$ are in or nearly in bound states. Therefore, one may expect the full spectral function to be the sum of three $\delta$ distributions at their pole positions. However, the strong mixing among them may enhance the phase shift contribution to the spectral function, especially when the enhancement is around a pole. The spectral function at $\mu_I=0.15$ GeV and $T=\mu_B=0$ is shown in Fig.\ref{fig8}. Considering the mixing effect on the scattering phase shift, we take here the absolute value of the spectral function in the pion superfluid. To focus on the dramatic change around the pole corresponding to the new mode $\sigma$ at $\omega=0.64$ GeV, we have neglected in Fig.\ref{fig8} the two $\delta$ distributions around $\omega=0$ and $0.25$ GeV corresponding to the Goldstone mode $\pi_+$ and $\pi_-$. The sudden dropping down of the scattering phase shift before the $\sigma$ pole spreads the expected narrow $\sigma$ spectrum and even splits the peak into two. This means that the pole approximation for the new mode $\sigma$ in the pion superfluid is no longer a good approximation, even if the the width shown in Fig.\ref{fig1} is very narrow.

We should point out that the spread of $\sigma$ spectrum in pion superfluid phase is independent of the regularization scheme. Applying the Pauli-Villars regularization scheme~\cite{maopv} with parameters $G=3.44$ GeV$^{-2}$, $\Lambda=1.127$ GeV, $m_0=0.005$ GeV, Fig.\ref{fig9} depicts the spectral function around the new mode $\sigma$ at $\mu_I=0.15$ GeV and $T=\mu_B=0$. The wide and two-peak spectrum looks similar as in the hard cutoff scheme (Fig.\ref{fig8}), and it is resulted from the mixing among the mesons in pion superfluid phase. The only difference is the location of the peak, which is around $m_\sigma=0.48$ GeV (the mass of new $\sigma$ mode) in Pauli-Villars regularization scheme and around $m_\sigma=0.64$ GeV in hard cutoff scheme.

\section {Conclusion}
\label{s4}
The quark-antiquark scattering phase shift is controlled by not only the meson bound or resonant state which leads to a jump around the pole but also the quarks as a background which is continuous and independent of the meson properties. In pion superfluid phase, the poles of the meson propagator are no longer located at the ordinary meson masses, and the phase shift is dramatically modified by the meson mixing. As a result, the meson spectral function is strongly corrected when the change in the phase shift happens around a pole. In the NJL model, the pole approximation for the new mode $\sigma$ is no longer a good approximation due to its strong mixing with pions. The mixing enhances the width of the spectrum and even splits the peak into two in the pion superfluid.

\section*{Acknowledgement}
The work is supported by the NSFC under grant No. 11405122.

\end{document}